\documentclass[11pt,oneside,a4paper]{article}
\usepackage[left=1in,right=1in,top=1in,bottom=1in]{geometry}
\usepackage{amsmath,amssymb}

\usepackage{hyperref}

\usepackage{setspace}

\hypersetup{colorlinks=true, linkcolor=blue, citecolor=black, bookmarks=false, pdfstartview={FitH}}

\begin{document}

\title{\bf New modes for massive Dirac field \\
in higher dimensional Black Holes}

\author{
{\bf Ciprian A. Sporea}\\
\ \ \\
{\small West University of Timi\c soara, Faculty of Physics,}\\
{\small V. Parvan Ave. no. 4, 300223, Timi\c soara, Romania} \\
{\small e-mail 1: ciprian.sporea89@e-uvt.ro} \\
{\small e-mail 2: sporea\_89@yahoo.com}
}

\date{\small \today}

\maketitle

\begin{abstract}
In this letter we derive new modes for massive Dirac field in the background of a higher-dimensional Schwarzschild black hole. We use in our approach the Cartesian gauge defined in local frames. We work in the context of ADD-like theories, assuming that the Standard Model fields (fermions and bosons) live only on a $(3+1)$-dimensional brane and the extra dimensions of space can be large in size.
\ \ \\


{\bf PACS (2008):} 04.62.+v $\bullet$ 04.50.Gh \\

\end{abstract}

\section{Introduction}

One of the main goals of today's physics is to develop a unified picture of nature in which all four of the known forces (gravitational, electromagnetic, week and strong nuclear force ) will be unified. During the last decades many ideas were proposed for the unification of gravity with the rest of the forces. Many of these ideas are based on the possible existence of extra dimensions of space. Although at the beginning it was thought that all extra dimensions must be very tiny in size and compact, the work of Arkani-Hamed, Dimopoulos and Dvali(ADD)\cite{16,16a,16b} have showed that these (linear) extra dimensions could have submillimeter and even millimeter length scales. In these {\it Large Extra Dimensions} scenarios a solution for  solving the hierarchy problem was proposed\cite{16}, thus lowering the fundamental energy scale of gravity (Plank scale) up to the order of TeV.

In the ADD brane-world scenario all the Standard Models particles (fermions and bosons) are constrained to live only on a $(3+1)$-dimensional {\it brane} (our observable 4-D Universe), while in the {\it bulk} only gravity can propagate and possibly other unknown scalar fields. Thus one possible way for detecting the extra dimensions is to study how gravity affects the other fields that live on the brane. This can be done, for example, by studying higher dimensional black holes. The process of higher-dimensional black holes creation at TeV scale energies was in the past decade the subject of many articles\cite{17,17a,19,19a,20,20a}$^{,}$\cite{21,21a,21b,21c,21d,21e,21f} . It is assumed that once such objects are formed they will quickly evaporate through Hawking radiation\cite{18}. This evaporation process of a higher-dimensional black hole was also the subject of many studies in the literature\cite{22,22a,22b,22c,22d,23,23a,23b,23c}.

In the vast majority of these studies the field equations, that describe the motion of the particles in the $(3+1)D$ brane induced gravitational background of the higher dimensional black hole, where studied manly for massless particles \cite{22,22a,22b,22c,22d}. One of the reasons for that is that these studies are using the Newman-Penrose\cite{24} formalism in which the massless field equations may be treated more easily (for the massive case in this formalism see Ref. \cite{15a}). However, in order to treat the massive fields (especially the Dirac field), a more standard approach, for dealing with the spin of the field, proves to be more efficient. As a consequence, the massive fields in the context of higher-dimensional black holes have been so far less studied in the literature.

In this short letter we continue the work of Ref.\cite{14,15,15a} by solving the massive Dirac equation in the background of a Schwarzschild $(4+n)$ higher-dimensional black hole, using a different gauge technique than the ones used there. We chose to work in the so called Cartesian gauge proposed some time ago in Ref.\cite{1,1a}. The advantage of this gauge is that the separation of the angular variables can be done in a very similar way as in the central problems from flat space-times. Thus the angular  part of the massive Dirac equation is completely solved in terms of the usual 4-dimensional angular spinors \cite{4,26}. For the remaining radial problem we will use an approximation technique for finding analytical solutions in the near horizon zone, respectively in the far field zone.

We would like to underline that the solutions found by us in the near horizon zone form a new set of modes that have never been discussed elsewhere in the literature before. These modes can be expressed as a combination of Whittaker $M$ and $W$ functions and besides other relevant physical constants (like the black hole and fermion masses, angular quantum number $\kappa$, etc) they also contain the dependence on the number $n$ of extra space dimensions. Regarding the far field modes we have found that they are a combination of Bessel functions of the first ($J$) and second kind ($Y$). We have thus recovered in the far field zone similar solutions to those reported in ref\cite{14,15}. To our best knowledge all the solutions to the Dirac equation, found in the literature, in the far field zone, for the fermions located on the {\it brane}, are independent of the number of extra-dimensions. Our solutions are also included in this category. This is no surprise because an observer placed in the far field zone will see the black hole as a 4-dimensional object due to the compactification of the extra dimensions.

Starting form the above mentioned new set of modes, one can in principle use the matching technique to combine the near horizon ones to those from the far field zone in order to calculate analytical expressions for the massive fermion absorbtion probability and most importantly for calculating the low-energy Hawking radiation (and luminosity) emitted during the Schwarzschild phase\cite{17,7,9} in the life of a higher dimensional black hole.

In the next section (2) we present some key features of the Castesian gauge that we are using to study the massive fermion fields.  After deriving, in the first part of section 3, the general equation for a massive Dirac field constrained to move on the $(3+1)$ brane projected background of a $(4+n)$ higher-dimensional black hole, we proceed with our investigation by studying the near horizon (section 3.1) and far-field (section 3.2) behavior of spin $1/2$ particles. In the last section (4) we present our conclusions derived from this research.

\section{Preliminaries}

In this section we will briefly review, following Ref\cite{1,1a} , how one can write down the Dirac equation in central charts using the Cartesian gauge \cite{1} which explicitly points out the central symmetry of the manifold. The Dirac equation in curved space-times can be defined in a local frame chart as \cite{1,3}
\begin{equation}\label{P1}
 i\gamma^{\hat\alpha}D_{\hat\alpha}\psi - m\psi=0\,,
\end{equation}
taking the explicit form \cite{1,2}
\begin{equation}\label{P2}
i\gamma^{\hat\alpha}e_{\hat\alpha}^{\mu}\partial_{\mu}\psi - m\psi
+ \frac{i}{2} \frac{1}{\sqrt{-g}}\partial_{\mu}(\sqrt{-g}e_{\hat\alpha}^{\mu})
\gamma^{\hat\alpha}\psi
-\frac{1}{4}
\{\gamma^{\hat\alpha}, S^{\hat\beta \cdot}_{\cdot \hat\gamma} \}
\hat\Gamma^{\hat\gamma}_{\hat\alpha \hat\beta}\psi =0\,,
\end{equation}

Both natural indices, $\alpha,..,\mu, \nu,...$ and local indices $\hat\alpha,..,\hat\mu,...$ take the same values $0,1,2,3$ and $S^{\hat\beta \cdot}_{\cdot \hat\gamma}$ stands for the generators of the spinor representation of the $SL(2,C)$ group. The line element can be expressed in terms of the tetrads $e_{\hat\alpha}$, $\hat e^{\hat\alpha}$ and the 1-forms $\omega^{\hat\mu}=\hat e^{\hat\mu}_{\nu}dx^{\nu}$ as $ds^2=\eta_{\hat\alpha\hat\beta}\omega^{\hat\alpha}\omega^{\hat\beta}=g_{\mu\nu}dx^{\mu}dx^{\nu}$. Using the tretrads defined in ref.\cite{1} one can show that the explicit form of the line element in a central static chart with spherical coordinates $(t, r, \theta, \phi)$ depends on three arbitrary functions $u(r)$, $v(r)$ and $w(r)$ and has the following expression\cite{1}
\begin{equation}\label{P3}
ds^{2}=w(r)^{2}\left[dt^{2}-\frac{dr^{2}}{u(r)^2}-
\frac{r^2}{v(r)^2}(d\theta^{2}+\sin^{2}\theta d\phi^{2})\right]\,.
\end{equation}
One of the main advantages of working with the Cartesian gauge is that the Dirac equation can be reduced to a much simpler form with the help of the transformation $\psi\to vw^{-\frac{3}{2}}\psi$. The new equation\cite{1} allows us to separate the spherical variables in the same manner done for the central problems in flat spacetimes. The positive frequency solutions of energy $E$ to this equation can then be written as
\begin{equation}\label{P4}
\Psi_{E,\kappa,m_{j}}(t,r,\theta,\phi)=\frac{v(r)}{rw(r)^{3/2}}[f^{+}_{E,\kappa}(r)\Phi^{+}_{m_{j},\kappa}(\theta,\phi)
+f^{-}_{E,\kappa}(r)\Phi^{-}_{m_{j},\kappa}(\theta,\phi)]e^{-iEt}\,,
\end{equation}
where $\Phi^{\pm}_{m_{j}, \kappa}$ are the usual four-component angular spinors \cite{4,5} and $f^{\pm}_{E,\kappa}(r)$ represents the radial wave functions. One can show that these functions satisfy a system of radial equations, written in compact form as

\begin{equation}\label{P5}
\underbrace{\left(\begin{array}{cc}
    m\,w(r)& -u(r)\frac{\textstyle d}{\textstyle dr}+\kappa\frac{\textstyle v(r)}
{\textstyle r}\\
&\\
  u(r)\frac{\textstyle d}{\textstyle dr}+\kappa\frac{\textstyle v(r)}
{\textstyle r}& -m\,w(r)
\end{array}\right)}_{the\,\, radial\,\, Hamiltonian\,\, H_r}
\left(\begin{array}{cc}
    f^{+}_{E,\kappa}(r)\\
&\\
  f^{-}_{E,\kappa}(r)
\end{array}\right)=E \left(\begin{array}{cc}
    f^{+}_{E,\kappa}(r)\\
&\\
  f^{-}_{E,\kappa}(r)
\end{array}\right)
\end{equation}

In what follows we will solve this eigenvalue problem for massive fermions constrained to move only on the brane.

\section{The Dirac equation in higher dimensional Black Holes}

In theories with extra dimensions of space one can discuss various problems. Among the most important ones is the study of black hole production\cite{6,7,17,17a,19,19a,20,20a,21,21a,21b,21c,21d,21e,21f} and the proprieties such objects may have  \cite{8}. Once produced these mini black holes will evaporate through Hawking radiation almost immediately. During this process the black hole will pass through several phases\cite{17,7,9}: (1) the {\it balding phase} during which the black hole emits mainly gravitational radiation; (2) the {\it spin-down phase}, in which the initial angular momentum of the black hole it is lost via Hawking radiation and superradiance; (3) the {\it Schwarzschild phase} when the black hole becomes spherically-symmetric and it's mass will decrease as the Hawking temperature increases due to the emission of radiation; (4) the {\it Plank phase} when the mass of the black hole is close to the Planck scale, phase for which we need a quantum theory of gravitation.

In our study we will concentrate on the Schwarzschild phase in the context of ADD-like theories\cite{16,16a,16b}. The purpose of this section is to study the brane fermion fields by solving the massive Dirac equation in the 4-dimensional induced background of the bulk black hole metric. For this we will apply the procedure described in section 2. We would like to point out that, up to our knowledge, this is the first time when one tries to study the brane fermions using the cartesian gauge\cite{1} for solving the Dirac equation.

A key feature of the ADD-like theories is the assumption that the Standard Model fields (fermions and gauge bosons) live only on a 4-dimensional brane, thus they will propagate in the gravitational background described by the $4D$ induced metric:
\begin{equation}\label{BH1}
ds^{2}=f(r)\,dt^{2}-\frac{dr^{2}}{f(r)}- r^{2} (d\theta^{2}+\sin^{2}\theta~d\phi^{2})\,,
\end{equation}
obtained from the higher-dimensional (4+n)-Schwarzschild black hole which has the line element\cite{6,6a}
\begin{equation}\label{BH2}
ds^{2}=f(r)\,dt^{2}-\frac{dr^{2}}{f(r)}- r^{2}d\Omega^2_{2+n}
\end{equation}
The induced metric (\ref{BH1}) is obtained form (\ref{BH2}) by setting $\theta_i=\pi/2$ ($i\ge2$) in the angular part of the metric $d\Omega^2_{2+n}$ given by
\begin{equation}\label{BH3}
d\Omega^2_{2+n}=d\theta^{2}_{n+1}+\sin^{2}\theta_{n+1} \left \{ d\theta^{2}_n+\sin^{2}\theta_n \left[ ... + \sin^2\theta_2 (d\theta^{2}_1+\sin^{2}\theta_1~d\phi^2)... \right] \right\}
\end{equation}
The function $f(r)$, both for the general metric (\ref{BH2}) and the induced one (\ref{BH1}), has the following expression
\begin{equation}\label{BH4}
f(r)=1-\left(\frac{r_{H}}{r}\right)^{n+1}
\end{equation}
where $n$ gives the number of space-like extra dimensions and $r_H$ represents the horizon radius, obtained by applying the Gauss law in a $(4+n)$-dimensional spacetime. Now $r_H$ has a more complicated expression (than the usual Schwarzschild, $r_0=2M_{BH}G$, radius) which depends on the number of extra dimension $n$, the mass of the black hole $M_{BH}$ and the fundamental higher-dimensional Plank mass $M_{*}$ (related to the standard Plank mass by: $M^2_P\approx M^{2+n}_{*}R^n $, with $R$ the size of the extra dimensions) given by\cite{6}
\begin{equation}\label{BH5}
r_H=\frac{1}{\pi M_{*}}\left( \frac{M_{BH}}{M_{*}} \right)^{\frac{1}{n+1}} \left( \frac{8\Gamma(\frac{n+3}{2})}{n+2} \right)^{\frac{1}{n+1}}
\end{equation}

Comparing (\ref{P3}) with (\ref{BH1}) we can identify the functions
\begin{equation}\label{BH6}
u(r)=f(r)=1-\left(\frac{r_{H}}{r}\right)^{n+1}, \qquad w(r)=v(r)=\sqrt{f(r)}=\sqrt{1-\left(\frac{r_{H}}{r}\right)^{n+1}}
\end{equation}
which enter into the radial Hamiltonian $H_r$ given in (\ref{P5}). As done in Ref\cite{10}. (for the standard 4D Schwarzschild black hole) we introduce the Novikov dimensionless coordinate\cite{11}
\begin{equation}\label{BH7}
x=\sqrt{\frac{r}{r_{H}}-1}\,\in\,(0,\infty)\,,
\end{equation}
with the help of which the exact radial eigenvalue problem $H_r{\cal F}=E{\cal F}$ in the space of two-component vectors ${\cal F}=(f^{+}, f^{-})^{T}$, can be rewritten as follows
\begin{equation}\label{BH8}
\left(\begin{array}{cc}
 A
& \quad B\\
&\\
C & \quad D
\end{array}\right)\,\left(
\begin{array}{c}
f^{+}(x)\\
\\
f^{-}(x)
\end{array}\right)=0\,,
\end{equation}
where we introduced the notations $\mu=r_{H}m\,,\epsilon=r_{H}E$ and
\begin{equation}\label{BH9}
\begin{split}
&A=\frac{\mu}{\sqrt{f}}-\frac{\epsilon}{f}\,, \qquad \qquad \qquad \,\,\,\,\,\,\,\,\,\, B=-\frac{1}{2x}\frac{d}{dx}+\frac{\kappa}{(x^2+1)\sqrt{f}} \\
&C=\frac{1}{2x}\frac{d}{dx}+\frac{\kappa}{(x^2+1)\sqrt{f}}\,, \qquad \,\,\,D=-\frac{\mu}{\sqrt{f}}-\frac{\epsilon}{f} \\
\end{split}
\end{equation}

In the next subsections we will solve equation (\ref{BH8}) by searching for analytical solutions in the vicinity of the black hole's horizon, respectively far away from it.

\subsection{Solutions in the near-horizon zone}

For finding the modes in the vicinity of the black hole we first need to approximate the operators given in (\ref{BH9}) by expanding them for small values of $x$. This can be done using a Taylor expansion for $x\approx 0$, obtaining the following expressions:
\begin{equation}\label{nh1}
\begin{split}
&\frac{1}{f}\approx \frac{1}{x}\frac{1}{n+1} \left [ \frac{1}{x} + \frac{n+2}{2}x + \frac{n(n+2)}{12}x^3+O(x^5) \right] \\
&\frac{1}{\sqrt{f}}\approx \frac{1}{x}\sqrt{\frac{1}{n+1}} \left [ 1+ \frac{n+2}{4}x^2 + \frac{n^2-4n-12}{96}x^4 +O(x^6)\right] \\
&\frac{1}{(x^2+1)\sqrt{f}}\approx \frac{1}{x}\sqrt{\frac{1}{n+1}} \left [1+ \frac{n-2}{4}x^2 +\frac{n^2-28n+36}{96}x^4+ O(x^6)\right]
\end{split}
\end{equation}
Keeping only the relevant terms (i.e. those proportional to $x$, $1/x$, respectively any constant terms) and neglecting the ones in $O(x^2)$ the system given in (\ref{BH8}) can be rewritten as
\begin{equation}\label{nh2}
\begin{split}
&  \underbrace{ \left [\frac{\mu}{\sqrt{n+1}}-\frac{\epsilon}{n+1}\left( \frac{1}{x}+\frac{n+2}{2}x \right)\right ]}_{\mathcal{A}}  f^{+}_{nh}(x) -
\underbrace{\left( \frac{1}{2}\frac{d}{dx}-\frac{\kappa}{\sqrt{n+1}}\right)}_{\mathcal{B}}f^{-}_{nh}(x)=0 \\
& \underbrace{\left( \frac{1}{2}\frac{d}{dx}+\frac{\kappa}{\sqrt{n+1}}\right)}_{\mathcal{C}}f^{+}_{nh}(x) -
\underbrace{\left [ \frac{\mu}{\sqrt{n+1}}+\frac{\epsilon}{n+1}\left( \frac{1}{x}+\frac{n+2}{2}x \right) \right ]}_{\mathcal{D}} f^{-}_{nh}(x)=0
\end{split}
\end{equation}
As it is, the above pair (\ref{nh2}) of differential equations can not be solved analytically. However, if we perform the transformation\cite{10}
\begin{equation}\label{nh3}
\left(\begin{array}{c}
f^{+}_{nh}(x)\\
\\
f^{-}_{nh}(x)
\end{array}\right)
=
\left(\begin{array}{cc}
 1
& \quad -i\\
&\\
-i & \quad 1
\end{array}\right)\,\left(
\begin{array}{c}
\hat f^{+}_{nh}(x)\\
\\
\hat f^{-}_{nh}(x)
\end{array}\right)\,,
\end{equation}
we obtain a new pair of differential equations for the new radial wave functions $(\hat f^{+}_{nh},\,\hat f^{-}_{nh})$ :
\begin{equation}\label{nh4}
\left [ (\mathcal{A}\pm \mathcal{C})\mp i(\mathcal{B}\pm \mathcal{D}) \right ]\hat f^{\pm}_{nh}(x) + \left [ (\mathcal{B}\pm \mathcal{D})\mp i(\mathcal{A}\pm \mathcal{C}) \right ]\hat f^{\mp}_{nh}(x) = 0
\end{equation}
The analytical solutions of (\ref{nh4}) can be found in terms of Whittaker $M$ and $W$ functions\cite{12} as
\begin{equation}\label{nh5}
\begin{split}
&\hat f^{+}_{nh}(x) = \frac{1}{\sqrt{x}}C_1 M_{\alpha,\beta}(i\gamma x^2)+ \frac{1}{\sqrt{x}}C_2 W_{\alpha,\beta}(i\gamma x^2)\\
&\hat f^{-}_{nh}(x) =\left(\frac{x}{2}p + \frac{1}{x}q \right) \left[ \frac{1}{\sqrt{x}}C_1 M_{\alpha,\beta}(i\gamma x^2)+ \frac{1}{\sqrt{x}}C_2 W_{\alpha,\beta}(i\gamma x^2) \right] \\
&- \frac{1}{x}\frac{i\sqrt{n+1}}{\kappa-i\mu}\left[ \left( \frac{1}{2} +\alpha+\beta \right)\frac{1}{\sqrt{x}}C_1 M_{\alpha+1,\beta}(i\gamma x^2) -  \frac{1}{\sqrt{x}}C_2 W_{\alpha+1,\beta}(i\gamma x^2)\right]\,,
\end{split}
\end{equation}
where we have introduced the following notations
\begin{equation}\label{nh6}
\begin{split}
&\alpha=\frac{1}{4}-i\left[\frac{\epsilon}{n+1}-\frac{\mu^2+\kappa^2}{\epsilon(n+2)} \right] \\
&\beta=\frac{1}{4}+i\frac{\epsilon}{n+1}, \qquad\gamma=\frac{\epsilon}{n+1}(n+2) \\
&p=\frac{\gamma(n+1)+\epsilon(n+2)}{\sqrt{n+1}(\kappa-i\mu)}, \qquad q=\frac{\epsilon+i(n+1)(\frac{1}{4}+\frac{\alpha}{\epsilon})}{\sqrt{n+1}(\kappa-i\mu)}
\end{split}
\end{equation}
while $C_1$ and $C_2$ are two arbitrary constants. These new set of modes are one of the main results of our research.

\subsection{Solutions in the far-field zone}

Let us now turn our attention to the far-field zone for which $r\gg r_H$. This can be done by taking the limit $x\rightarrow \infty$ in the relations (\ref{BH9}). By doing so we have found that there is a difference between the case $n=0$ (no extra dimensions) studied in Ref.\cite{10} and the case when one or more extra dimensions are present ($n\ge1$). Practically we have found that the Dirac equation has different particular solutions for the two cases. The solutions found in Ref.\cite{10} for $n=0$ are a combination of Whittaker M and W functions, while our solutions for $n\ge1$ can be expressed in terms of Bessel functions.

In order to obtain the system of differential equations resulted from (\ref{BH8}) for the far-field zone, we first need to calculate the following expansions  in the asymptotic region of the black hole:
\begin{equation}\label{ff1}
\begin{split}
&\frac{1}{f}\approx \frac{1}{x} \left [x+ \frac{1}{x^{2n+1}} - \frac{n+1}{x^{2n+3}} + O(1/x^{2n+5})\right] \\
&\frac{1}{\sqrt{f}}\approx \frac{1}{x} \left [x+ \frac{1}{2x^{2n+1}} + O(1/x^{2n+3})\right] \\
&\frac{1}{(x^2+1)\sqrt{f}}\approx \frac{1}{x} \left [\frac{1}{x}- \frac{1}{x^3} + O(1/x^5)\right]
\end{split}
\end{equation}
which are in fact the Taylor series with respect to $1/x$ of the operators given in (\ref{BH9}). Keeping only the terms proportional to $x$, respectively $1/x$ and any constant terms, neglecting thus the terms of the order $O(1/x^2)$, we arrive at the following pair of equations for the radial wave functions $(f^{+}_a, f^{-}_a)$:
\begin{equation}\label{ff2}
\begin{split}
&\left( \frac{1}{2}\frac{d}{dx}+\frac{\kappa}{x} \right)f^{+}_a(x) - x(\epsilon+\mu) f^{-}_a(x)=0 \\
& x(\epsilon-\mu) f^{+}_a(x) + \left( \frac{1}{2}\frac{d}{dx}-\frac{\kappa}{x} \right)f^{-}_a(x)=0
\end{split}
\end{equation}
which is a system of two Bessel differential equations\cite{12} with the general solutions (in the asymptotic zone) given by
\begin{equation}\label{ff3}
\begin{split}
& f^{+}_{a}(x) = \sqrt{\epsilon+\mu} \left[ a\,x J_{\kappa+\frac{1}{2}}(\nu x^2)+b\,x\, Y_{\kappa+\frac{1}{2}}(\nu x^2) \right] \\
&f^{-}_{a}(x) = \sqrt{\epsilon-\mu} \left[ a\,x J_{\kappa-\frac{1}{2}}(\nu x^2)+b\,x\, Y_{\kappa-\frac{1}{2}}(\nu x^2) \right]
\end{split}
\end{equation}
where we denoted by $\nu=\sqrt{\epsilon^2-\mu^2}$, while $a$ and $b$ are two arbitrary coefficients.

Because our modes are a combination of Bessel functions of the first and second kind, one can not talk about the existence of discrete energy levels like in the case of black holes with no extra dimensions\cite{10,13}. This is again a difference between the classical Schwarzschild black holes and their higher-dimensional counterparts. We think that a full investigation of the exact radial problem (\ref{BH8}) using numerical methods\cite{27,28,13} for solving the Dirac equation (in the Cartesian gauge) may reveal the existence of discrete energy levels for fermions moving in the background of a higher-dimensional black hole.

\section{Conclusions}

In this study we investigated the brane behaviour of massive fermions in the vicinity and far away from a $(4+n)$-dimensional black hole. This was done by finding new analytical solutions to the Dirac equation in the near horizon, respectively in the far-field zone of the black hole. We have thus proved once more the advantages of using the Cartesian gauge for investigating the (massive) Dirac equation in different curved backgrounds.

This research can be continued in a number of ways. One of these is the full investigation of the radial problem with the help of numerical techniques in order to find a complete (numerical) solution. It would be also interesting to derive the Hawking radiation of massive fermions using our new modes given here, and to see to what extent this radiation will differ from the one obtained in other studies found in the literature.

\section*{Acknowledgments}

The author would like to thank Professors I.I. Cotaescu and C. Crucean for reading the manuscript and for useful remarks and discussions that helped to improve this work.

C.A. Sporea was supported by the strategic grant
POSDRU/159/1.5/S/137750, Project “Doctoral and Postdoctoral programs support for increased competitiveness
in Exact Sciences research” cofinanced by the European Social Found within the Sectorial Operational Program Human
Resources Development 2007-2013.

\end{document}